\documentclass[dvipdfmx]{PoS}
\usepackage{amsmath}
\usepackage{url}
\usepackage{here}

\makeatletter
\newcommand*{\textoverline}[1]{$\overline{\hbox{#1}}\m@th$}
\makeatother

\title{Mass and Axial current renormalization in the Schr\"{o}dinger functional scheme 
for the RG-improved gauge and the stout smeared $O(a)$-improved Wilson quark actions}

\ShortTitle{Mass and Axial current renormalization in the Schr\"{o}dinger functional scheme ...}

\author{PACS collaboration: \speaker{K.-I. Ishikawa}$^{a,d}$,
N.~Ishizuka$^{b,c}$,
Y.~Kuramashi$^{b,c,d}$,
Y.~Nakamura$^{d}$,
Y.~Namekawa$^{b}$,
Y.~Taniguchi$^{c}$,
N.~Ukita$^{b}$,
T.~Yamazaki$^{b,c,d}$,
T.~Yoshi\'{e}$^{b,c}$\\
$^{a}$Graduate School of Science, Hiroshima University, Higashi-Hiroshima, Hiroshima 739-8526, Japan\\
$^{b}$Center for Computational Sciences, University of Tsukuba, Tsukuba, Ibaraki 305-8577, Japan\\
$^{c}$Graduate School of Pure and Applied Sciences, University of Tsukuba, Tsukuba, Ibaraki 305-8571, Japan\\
$^{d}$RIKEN Advanced Institute for Computational Science, Kobe, Hyogo 650-0047, Japan\\
E-mail: \email{ishikawa@theo.phys.sci.hiroshima-u.ac.jp}}

\abstract{
 {\normalsize        %
 \vspace*{-33.5em}   %
 \begin{flushright}  
 HUPD-1509, UTHEP-680, UTCCS-P-83%
 \end{flushright}    %
 \vspace*{31em}\     %
 }                   %
We present the quark mass and axial current renormalization factors for 
the RG-improved Iwasaki gauge action and 
three flavors of the stout smeared $O(a)$-improved Wilson quark action.
We employ $\alpha=0.1$ and $n_{\mathrm{step}}=6$ for the stout link smearing 
parameters and all links in 
the quark action are replaced with the smeared links.
Using the Schr\"{o}dinger functional scheme we evaluate the renormalization 
factors at $\beta=1.82$ where large scale simulations
are being carried out.}

\FullConference{The 33rd International Symposium on Lattice Field Theory\\
                14 -18 July  2015\\
                Kobe International Conference Center, Kobe, Japan}

\begin{document}

\section{Introduction}
We are accumulating the configurations near the physical masses of 
up-down (degenerated) and strange quarks ($N_f=2+1$) on a $96^4$ lattice 
with the lattice cutoff $a^{-1}\sim 2.3$ [GeV] under the project of 
HPCI (High Performance Computing Infrastructure) 
Strategic Programs for Innovative Research (SPIRE) Field 5,
``The origin of matter and the universe''~\cite{HPCIF5}. 
The properties of these ensembles have to be determined via the detailed analysis on 
the physical quantities such as the light hadron spectrum, quark masses, and hadronic observables,
extracted from the ensembles.

In this poster, we report on the mass and axial current renormalization
factors determined with the Schr\"{o}dinger functional (SF) scheme 
for the RG-improved Iwasaki gluon action and three flavors of 
the stout smeared $O(a)$-improved Wilson quark action. 
We employ $\alpha=0.1$ and $n_{\mathrm{step}}=6$ for the stout link smearing~\cite{Morningstar:2003gk} parameters, 
and all link variables contained in the $O(a)$-improved Wilson quark action are replaced 
with the stout smeared ones. The $O(a)$-improvement parameter $c_{\mathrm{SW}}$ 
has been determined nonperturbatively in Ref.~\cite{Taniguchi:2012kk} using the SF method.
We determine the renormalization factors at $\beta=1.82$ ($a^{-1}\sim 2.3$ [GeV])
where the simulations on the $96^4$ lattice with this lattice action are being carried out.

The renormalization constants for the axial and pseudo-scalar operators (and vector operator as a byproduct)
with the SF scheme are determined using the standard method described in 
Refs.~\cite{Luscher:1996jn,Della Morte:2005rd,Bulava:2015vja,Hoffmann:2005cz,Aoki:2010wm}.
The temporal and spatial lattice sizes are finite at $T=a N_T$ and $L = a N_S$,
and a Dirichlet boundary condition in the temporal direction is imposed to define the SF scheme.
The boundary gauge fields at $n_4=0$ and $N_T$ are kept fixed by the Dirichlet boundary 
condition and the same boundary condition is applied on the smeared gauge field 
during the stout smearing steps.
The up, down and strange quark masses are degenerate and tuned to vanish in determining 
the renormalization constants.
The HMC (two-flavor part) and RHMC (single-flavor part) algorithms 
with the SF boundary condition are employed to generate the gauge configuration.
The action parameters are set to be $\beta=1.82$ and $c_{\mathrm{SW}}=1.11$~\cite{Taniguchi:2012kk}, and 
the boundary parameters are to be $c_t^{P}=1$ ($c_t^{R}=3/2$) for the plaquette (rectangular) 
term for the gauge action and $\tilde{c}_t=1$ for the quark action. 

\section{Axial, vector, and pseudo-scalar operator renormalization constants}
The operators to be renormalized are defined by
\begin{align}
A_4^a(x)&=\overline{q}(x)\gamma_4\gamma_5T^a q(x),&
V_4^a(x)&=\overline{q}(x)\gamma_4T^a q(x),&
P^a(x)  &=\overline{q}(x)\gamma_5T^a q(x).
\end{align}
As it was observed that the nonperturbative $O(a)$-mixing correction to 
the axial current was consistent with zero~\cite{Taniguchi:2012kk}, we assume 
a negligible $O(a)$-mixing and employ the unimproved current operators in determining the renormalization factors.
The correlation functions used for setting the renormalization conditions are 
\begin{align}
f_{XY}(t,s) &= - \dfrac{2}{N_f^2(N_f^2-1)}\sum_{\vec{x},\vec{y}}
f^{abc}f^{cde}
\langle O'^d X^a(\vec{x},t)Y^b(\vec{y},s)O^e\rangle,
\label{eq:FXY}
\\
f_X(t) &= -\dfrac{1}{N_f^2-1}\sum_{\vec{x}}\langle X^a(\vec{x},t)O^a\rangle,
\label{eq:FX}
\\
f_1 &= - \dfrac{1}{N_f^2-1}\langle O'^a O^a\rangle,
\label{eq:F1}
\\
f_V(t) &= \dfrac{1}{N_f(N_f^2-1)}\sum_{\vec{x}}if^{abc}\langle O'^a V_4^b(\vec{x},t)O^c\rangle,
\label{eq:FV}
\end{align}
where $f^{abc}$ is the structure constant of $SU(N_f)$.
The operator $X$ ($Y$) can be $A_4$ or $P$.
$O^a$ and $O'^a$ are boundary operators defined by
\begin{align}
 O^a  &= \dfrac{1}{(L/a)^3}\sum_{\vec{y},\vec{z}}\overline{\zeta}(\vec{y})\gamma_5 T^a\zeta(\vec{z}),
&O'^a &= \dfrac{1}{(L/a)^3}\sum_{\vec{y},\vec{z}}\overline{\zeta}'(\vec{y})\gamma_5 T^a\zeta'(\vec{z}),
\end{align} 
where $\zeta$ and $\zeta'$ are boundary quark fields located at $n_4=0$ and $n_4=N_T$ respectively.

Using the correlation functions (\ref{eq:FXY})-(\ref{eq:FV}), the renormalization factors are 
determined with
\begin{align}
  Z_A &= \left. \sqrt{\tilde{Z}_A(2T/3)}\right|_{m_q \to 0},
&
\tilde{Z}_A(t)&=
\dfrac{f_1}{n_A}\left[
f_{AA}(t,T/3)-2 m_{\mathrm{PCAC}}f_{PA}(t,T/3)
\right]^{-1},
\label{eq:ZA}
\\
  Z_V &= \left. \tilde{Z}_V(T/2)        \right|_{m_q \to 0},
&
\tilde{Z}_V(t)&=\dfrac{f_1}{n_V f_V(t)},
\label{eq:ZV}
\\
  Z_P &= \left. \tilde{Z}_P(T/2)        \right|_{m_q \to 0},&
\tilde{Z}_P(t)&=\dfrac{\sqrt{3 f_1}}{n_P f_P(t)},
\label{eq:ZP}
\end{align}
where $n_A$, $n_V$, and $n_P$ are normalization constants evaluated
at the tree-level so as to be $Z_A=1$, $Z_V=1$, and  $Z_P=1$.
To take the mass-less limit we employ 
an averaged PCAC mass for $Z_A$ and $Z_V$;
\begin{align}
am_q = am^{\mathrm{PCAC}}=
\dfrac{1}{3}\sum_{t=T/2-a}^{T/2+a}
\dfrac{af_A(t+a)-af_A(t-a)}{4f_P(t)},
\end{align}
while a non-averaged mass for $Z_P$;
\begin{align}
 am_q=am^{\mathrm{PCAC}} = \dfrac{af_A(T/2+a)-af_A(T/2-a)}{4f_P(T/2)}.
\label{eq:mPCACnonaved}
\end{align}

The simulation parameters are shown in Table~\ref{tab:param}. 
The phase angle $\theta$ is the parameter of the generalized periodic boundary condition 
in each spatial direction for the quark field. 
The boundary gauge fields are fixed to the identity matrix.  
The data are measured at every trajectory and 
the statistical errors are estimated with the jackknife method after 
blocking data with the size of 100 trajectories.
The simulations (A1S) and (A1L) are dedicated for $Z_A$ (and $Z_V$ as a byproduct), 
while (P4a) and (P4b) are for $Z_P$. 
The hopping parameter $\kappa=0.126110$ corresponds to the critical value $\kappa_c$
determined in Ref.~\cite{Taniguchi:2012kk}. Almost vanishing masses are realized in the (A1S) and (A1L) runs.  

\begin{figure}[t]
    \centering
    \includegraphics[scale=0.58]{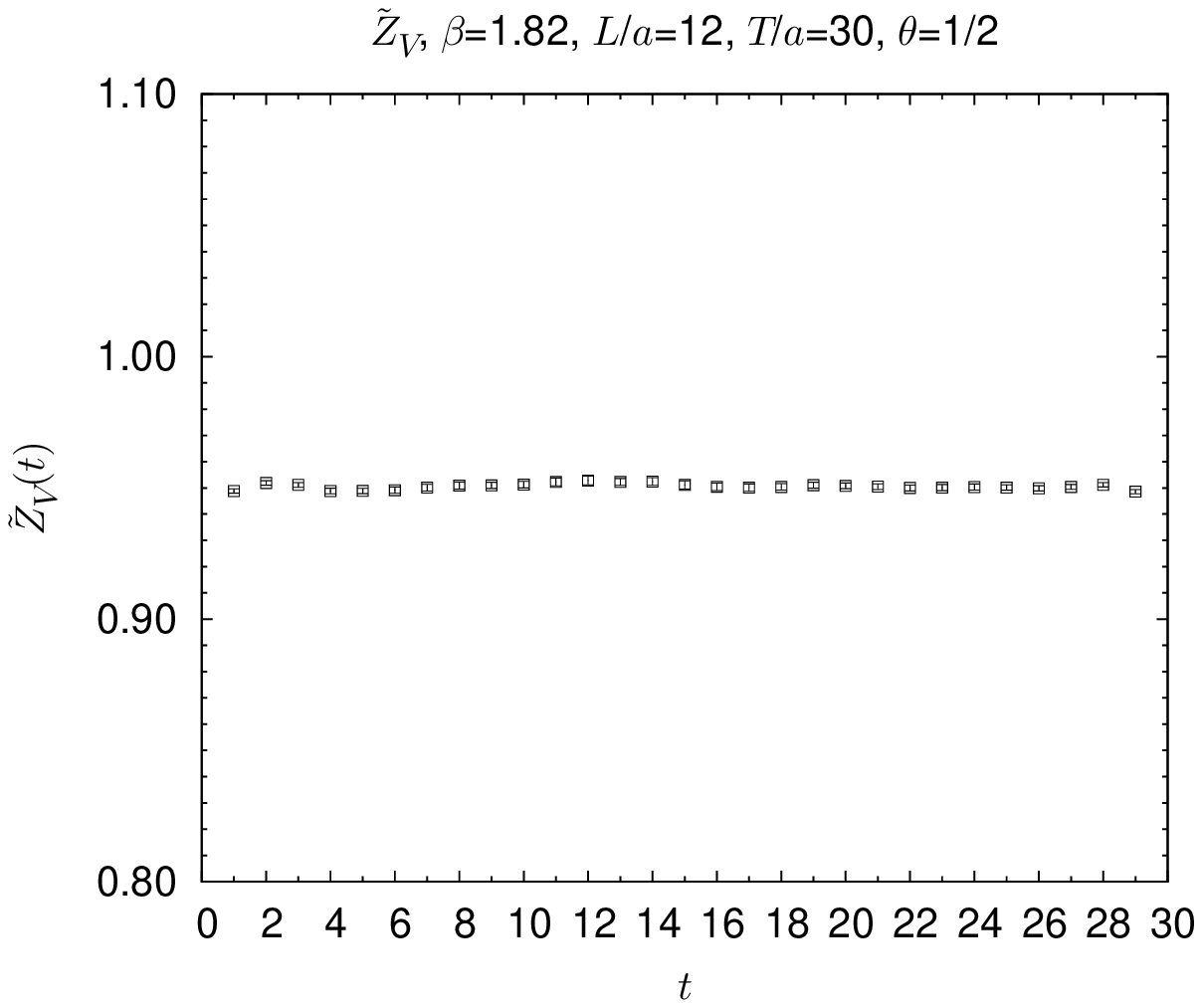}
\hfil
    \includegraphics[scale=0.58]{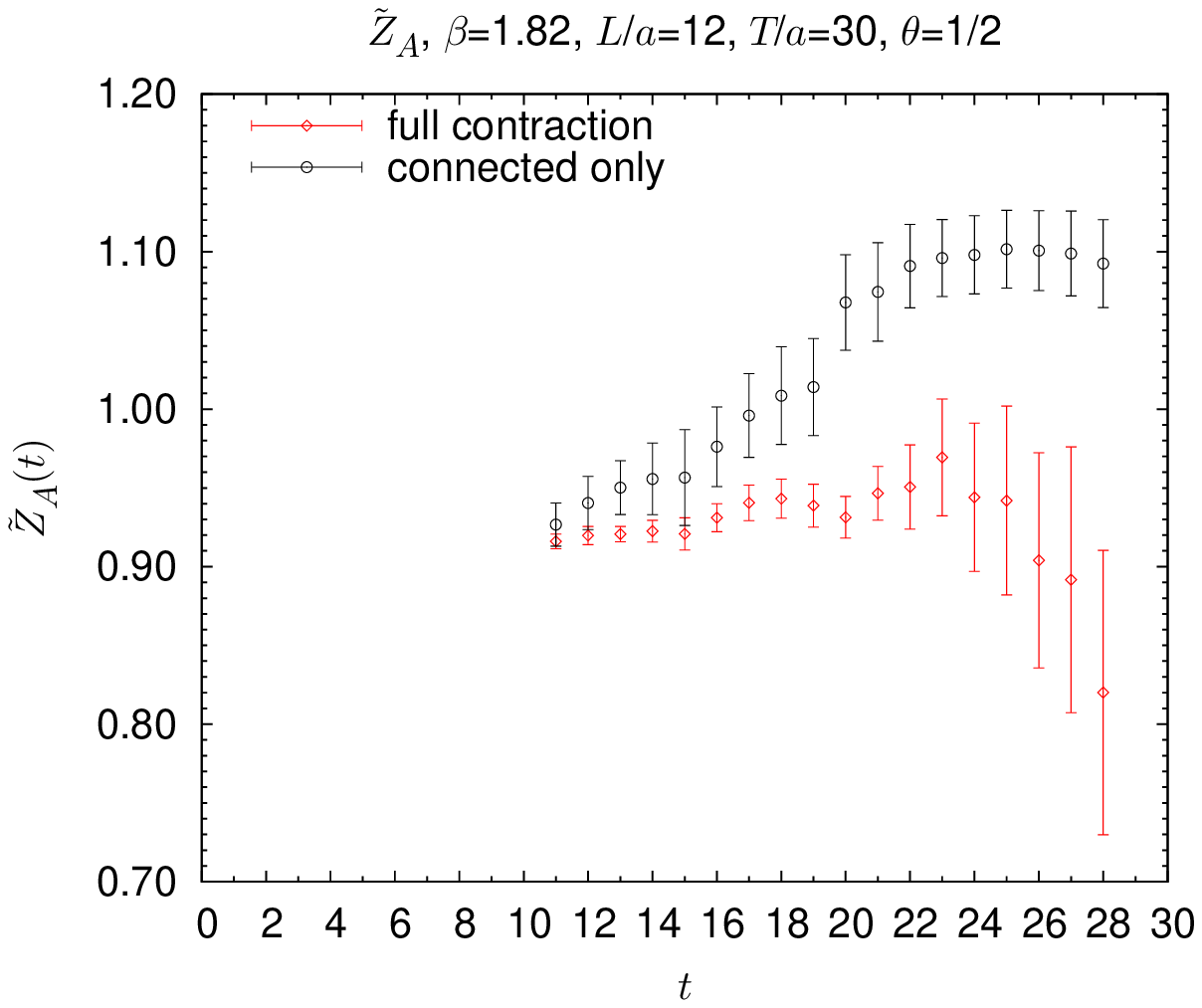}
    \caption{Time dependence of $\tilde{Z}_V(t)$ (left) and $\tilde{Z}_A(t)$ (right) on a $12^3\times 30$ lattice (A1L).}
    \label{fig:tildeZVA}
\end{figure}

The time dependence of $\tilde{Z}_A(t)$ and $\tilde{Z}_V(t)$ from (A1L) is shown in Figure~\ref{fig:tildeZVA}.
$\tilde{Z}_V(t)$ is almost time-independent, and 
a short plateau around $t=17-22$ ($\sim 2T/3$) is observed for $\tilde{Z}_A(t)$ 
when disconnected diagrams are included properly. 
A similar behavior is observed in (A1S).
The renormalization factors extracted with the definitions (\ref{eq:ZA})-(\ref{eq:ZP})
are tabulated in Table~\ref{tab:zfac}.
We assign the discrepancy of $Z_A$ and $Z_V$ between two runs, (A1S) and (A1L), to the systematic error.
We observe that $Z_A \simeq Z_V$ and $Z_A\simeq 1$. 
This could indicate a better chiral property of the stout smeared quark action we employed.

The renormalization factor for the pseudo-scalar operator depends on the renormalization scale.
The renormalization scale corresponds to the physical box size $L$ and
the scale is implicitly defined by the value of the renormalized coupling ($g_{\mathrm{SF}}^2(L)$)
in the SF scheme. To convert $Z^{\mathrm{SF}}_P(1/L)$ to the mass renormalization factor 
$Z^{\overline{\mathrm{MS}}}_m(\mu)$ at a reference scale $\mu$ in the \textoverline{MS} scheme, 
we need the RG evolution of $Z_P$ together with the running of the renormalized coupling constant $g^2(\mu)$
in both the SF and \textoverline{MS} schemes. 
The details of the mass renormalization factor are described in the next section.

\begin{table}[t]
    \centering
    \begin{tabular}{cccccc}\hline\hline
 Run  & $L/a, T/a$ & $\theta$ & $\kappa$ & traj. & HMC Acc. \\ \hline
(A1S) &  $8 , 18$ & 1/2 & 0.126110 &  20000  & 0.8811(80)   \\
(A1L) & $12 , 30$ & 1/2 & 0.126110 &  34700  & 0.9120(53)   \\\hline
(P4a) &  $4 ,  4$ & 1/2 & 0.126110 & 100000  & 0.8792(11)   \\
(P4b) &  $4 ,  4$ & 1/2 & 0.125120 &  80000  & 0.8787(15)   \\\hline\hline
    \end{tabular}
    \caption{Parameters and statistics for the renormalization factors.}
    \label{tab:param}
\end{table}
\begin{table}[t]
    \centering
    \begin{tabular}{ccccc}\hline\hline
 Run  & $am^{\mathrm{PCAC}}$ & $Z_V$ & $Z_A$     & $Z_P$ \\ \hline
(A1S) &  0.00041(61) &  0.9664(20)  & 0.9745(48) &  -    \\
(A1L) & -0.00080(33) &  0.95153(76) & 0.9650(68) &  -    \\\hline
(P4a) & -0.021859(94)&  - & - & 1.01317(43)\\ 
(P4b) &  0.013241(99)&  - & - & 1.00670(45)\\\hline\hline
    \end{tabular}
    \caption{PCAC masses and renormalization factors $Z_V$, $Z_A$, and $Z_P$.}
    \label{tab:zfac}
\end{table}

\section{Scale setting and mass renormalization}

The renormalization factor of the pseudo-scalar operator $Z_P$ is evaluated at $T=L=4a$ and $\beta=1.82$.
The  simulation parameters and results are shown 
in Tables~\ref{tab:param} and \ref{tab:zfac}. We have two simulations, (P4a) and (P4b).
The scale $L=4a$ is chosen so that the RG evolution by the step scaling of the coupling is available 
from the scale $L_{\mathrm{max}}=4a$.
The renormalized coupling in the SF scheme at $L=4a$ and $\beta=1.82$ is shown in Table~\ref{tab:GSF}, 
where the boundary condition defining the SF scheme coupling is imposed on the gauge field.
The PCAC mass in Table~\ref{tab:GSF} is defined by Eq.~(\ref{eq:mPCACnonaved}).
The step scaling evolution for the SF scheme coupling from $g_{\mathrm{SF}}^2(1/L_{\mathrm{max}})\sim 3.7-3.8$ 
is available in the continuum limit~\cite{Aoki:2009tf}.
The PCAC masses at $\kappa_c=0.126110$ is slightly off the vanishing point as seen in (P4a) and (G4a) runs.
The discrepancy between $\kappa=0.12512$ and $\kappa_c$ are assigned to 
the systematic error of $Z_m$.

Combining the RG evolutions for $Z_P$ and the coupling in both schemes,
we can extract the mass renormalization constant $Z^{\overline{\mathrm{MS}}}_m(g_0,\mu)$ 
in the \textoverline{MS} scheme~\cite{Aoki:2010wm,Aoki:2009tf,Capitani:1998mq,DellaMorte:2005kg} by
\begin{align}
Z_m^{\mathrm{\overline{MS}}}(g_0,\mu)
&= 
\left(\dfrac{m^{\mathrm{\overline{MS}}}(\mu)}{M^{\mathrm{RGI}}}\right)
\left(\dfrac{M^{\mathrm{RGI}}}{m^{\mathrm{SF}}(1/L_{\mathrm{max}})}\right)
\left(\dfrac{Z_{A}^{\mathrm{SF}}(g_0,a/L)}{Z_{P}^{\mathrm{SF}}(g_0,a/L_{\mathrm{max}})}\right),
\label{eq:DefZm}
\end{align}
where $L_{\mathrm{max}}=4a$ will be applied.
The mass ratios between the renormalization group invariant (RGI) mass 
$M^{\mathrm{RGI}}$ and the renormalized masses $m^{\overline{\mathrm{MS}}}$ 
(or $m^{\mathrm{SF}}$) are defined by
\begin{align}
\left(\dfrac{M^{\mathrm{RGI}}}{m^{\mathrm{\overline{MS}}}(\mu)}\right) 
&=\left.\left(2b_0 \bar{g}^2(\mu)\right)^{-d_0/(2b_0)}
\exp\left[-\int^{\bar{g}(\mu)}_0 
dg \left(\dfrac{\tau(g)}{\beta(g)}-\dfrac{d_0}{b_0 g}\right)\right]
\right|_{\overline{\mathrm{MS}}},
\label{eq:MSPevo}
\\
\left(\dfrac{M^{\mathrm{RGI}}}{m^{\mathrm{SF}}(1/L_{\mathrm{max}})}\right) 
&=
\left.
\left[\prod_{j=1}^{n}\sigma_{P}(u_j)\right]
\left(2b_0 \bar{g}^2(1/L_n)\right)^{-d_0/(2b_0)}
\exp\left[-\int^{\bar{g}(1/L_n)}_0
dg 
\left(\dfrac{\tau(g)}{\beta(g)}-\dfrac{d_0}{b_0 g}\right)\right]
\right|_{\mathrm{SF}}.
\label{eq:SFPevo}
\end{align}
$\bar{g}^2(\mu)$ is the renormalized coupling 
constant 
in the \textoverline{MS} scheme for Eq.~(\ref{eq:MSPevo}), and 
that in the SF scheme for Eq.~(\ref{eq:SFPevo}).
$\sigma_P(u)$ is the step scaling function for $Z_P$ in the SF scheme.
The argument $u_j$ is the renormalized coupling defined by 
 $u_j=g^2_{\mathrm{SF}}(2^j/L_{\mathrm{max}})$ which is evolved 
from $u_0=g^2_{\mathrm{SF}}(1/L_{\mathrm{max}})$ using the step 
scaling function $\sigma(u)$ for the coupling via $u_{j+1}=\sigma(u_j)$.

In order to evaluate the mass renormalization constant $Z^{\overline{\mathrm{MS}}}_m(g_0,\mu)$,
we employ $\sigma_P(u)$ from Ref.~\cite{Aoki:2010wm} and $\sigma(u)$ from Ref.~\cite{Aoki:2009tf}.
The number of steps $n$ is chosen to be 5 from which
we can evaluate the exponent of Eq.~(\ref{eq:SFPevo}) perturbatively.
The two-loop mass anomalous dimension $\tau(g)$~\cite{Sint:1998iq}
 and the three-loop beta function $\beta(g)$~\cite{Bode:1999sm} are 
used in the SF scheme, while 
the four-loop estimates for the mass anomalous dimension and beta function
are employed in the \textoverline{MS} scheme in evaluating the mass renormalization constant. 
In order to evaluate the mass renormalization constant at the reference scale $\mu= 2$ [GeV], 
we need the scale of $a^{-1}$ in physical unit at $\beta=1.82$.

\begin{table}[t]
    \centering
    \begin{tabular}{ccccc|cc}\hline\hline
 Run  &  $\theta$ & $\kappa$ & traj. & HMC Acc.   & $am^{\mathrm{PCAC}}$ & $g_{\mathrm{SF}}^2$\\\hline
(G4a) &  $\pi/5$ & 0.126110 & 70000  & 0.8787(14) &  -0.04550(26) &   3.662(17) \\
(G4b) &  $\pi/5$ & 0.125120 & 80000  & 0.8791(15) &   0.00042(25) &   3.776(16) 
\\\hline\hline
    \end{tabular}
    \caption{Parameters, statistics and the renormalized coupling $g_{\mathrm{SF}}^2$ at $\beta=1.82$ and $L/a=T/a=4$.}
    \label{tab:GSF}
\end{table}

\section{Results}
The axial and vector current renormalization factors at $\beta=1.82$ are evaluated as
\begin{align}
Z_A &= 0.9650(68)(95),
\label{eq:ZAVAL}
\\
Z_V &= 0.95153(76)(1487),
\label{eq:ZVVAL}
\end{align}
where the central values are from (A1L). The first error is the statistical one and the second 
is the systematic one which is evaluated from the discrepancy between the two runs.

The mass renormalization constant $Z_m^{\overline{\mathrm{MS}}}(g_0,\mu)$
at $\mu =2$ [GeV] and $\beta= 1.82$ is evaluated as
\begin{align}
Z_m^{\overline{\mathrm{MS}}}(g_0,\mu=2\mbox{[GeV]}) = 0.9950(111)(89),
\label{eq:ZMVAL}
\end{align}
where the central value is extracted by combining
the factors $Z_A$ from (A1L), $Z_P$ from (P4a), and $u_0$ from (G4a).
For the scale $a^{-1}$, we use the preliminary value $a^{-1}=2.332(18)$ [GeV] from Ref.~\cite{LAT15:UKITA}.
The first error is the statistical error and the second is the systematic one estimated 
from the discrepancy to the mass renormalization constant evaluated using (P4b) and (G4b).

\section{Summary}
In this poster, we presented the determination of 
the renormalization constants for the axial and vector currents and the quark mass
in the Schr\"{o}dinger functional scheme. The values in 
Eqs.~(\ref{eq:ZAVAL})-(\ref{eq:ZMVAL}) were obtained for the RG-improved Iwasaki gluon
and three flavors of the stout smeared quarks at $\beta=1.82$.
By applying these renormalization factors to the results from 
the simulation on the $96^4$ lattice, we can obtain the quark masses and 
the decay constants precisely~\cite{LAT15:UKITA}.

\section*{Acknowledgments}
The numerical computations were performed using a PC-cluster 
of HPCI Strategic Program Field 5 
and the HA-PACS GPU cluster system at the Center for Computational Sciences, 
University of Tsukuba.
This work was supported in part by 
the Grant-in-Aid for Scientific Research (Nos. 15K05068, 25800138) from the Japan Society 
for the Promotion of Science (JSPS),
and by MEXT and JICFuS as a priority issue (Elucidation of the fundamental laws and evolution of the universe) 
to be tackled by using Post ``K'' Computer.

\end{document}